\documentclass[fleqn,usenatbib]{mnras}

\usepackage{newtxtext,newtxmath}

\usepackage[T1]{fontenc}

\DeclareRobustCommand{\VAN}[3]{#2}
\let\VANthebibliography\thebibliography
\def\thebibliography{\DeclareRobustCommand{\VAN}[3]{##3}\VANthebibliography}

\usepackage{graphicx}	
\usepackage{amsmath}	
\usepackage{color}
\usepackage{amsmath}
\usepackage{multirow}
\usepackage{supertabular}
\usepackage{tabularx}
\usepackage{hyperref}
\hypersetup{
    colorlinks=true,
    linkcolor=blue,
    filecolor=magenta,      
    urlcolor=cyan,
    citecolor=blue,
}
\usepackage{xspace}

\newcommand\who{WhoSGlAd\xspace}
\newcommand\whoac{\textbf{Who}le \textbf{S}pectrum and \textbf{Gl}itches \textbf{Ad}justment\xspace}

\title[Second helium ionisation zone acoustic depth study]{Study with \who of the acoustic depth of the helium glitch across the seismic HR diagram and its impact on the inferred helium abundance}
\author[M.~Farnir et al.]{M.~Farnir$^1$
A. Valentino$^2$
M-A.~Dupret$^3$
A.-M. Broomhall$^1$
\\
$^1$Centre for Fusion, Space, and Astrophysics, Department of Physics, University of Warwick, Coventry, CV4 7AL, United Kingdom \thanks{e-mail: martin.farnir@warwick.ac.uk} \\
$^2$Center for Mathematical Plasma Astrophysics, KU Leuven, Leuven 3000, Belgium \\
$^3$Institut d’Astrophysique et G\'eophysique de l’Universit\'e de Li\`ege, All\'ee du 6 ao\^ut 17, 4000 Li\`ege, Belgium}
\date{Accepted XXX. Received YYY; in original form ZZZ}

\pubyear{2022}

\begin{document}
\label{firstpage}
\pagerange{\pageref{firstpage}--\pageref{lastpage}}
\maketitle

\begin{abstract}
    The acoustic glitches’ signature present in solar-like stars holds invaluable information. Indeed, it is caused by a sharp variation in the sound speed, therefore carrying localised information. One such glitch is the helium glitch caused by the hydrogen and first and second partial helium ionisation region, allowing us to constrain the surface helium abundance. However, the function adjusted to the glitch signature depends non-linearly on the acoustic depth at which it occurs, He. Retrieving the faint glitch signature and estimating $\tau_{\textrm{He}}$ are difficult but crucial tasks to accurately measure the glitch parameters and, ultimately, accurately infer the helium abundance.\\
    In the present paper, we aim at providing a way to estimate $\tau_{\textrm{He}}$ using precise seismic indicators, independent of stellar modelling. Consequently, we aim at improving the \who (\whoac) method by automatically providing a model independent measure of the glitch’s parameters.\\
    We compute the evolution of $T_{\textrm{He}}$, a dimensionless form of the acoustic depth, along a grid of models and adjust an empirical linear relation between $T_{\textrm{He}}$ and the mean large separation and frequency ratio as defined in \who. We further optimise over the value of this estimate to ensure the stability and accuracy of the approach.\\
    The proposed approach provides an excellent estimate of the acoustic depth and allows us to swiftly retrieve the glitch signature of observed spectra. We demonstrate that the we can accurately model the helium abundance of four Kepler targets by comparing model (both versions of \who) and literature values.
\end{abstract}

\section{Introduction}
In recent years the detection and precise measurement of stellar oscillation modes has propelled forward the field of stellar physics. Indeed, thanks to recent space-based surveys \citep[such as CoRoT and Kepler][]{2009IAUS..253...71B,2010AAS...21510101B} providing data of unprecedented quality, asteroseismology, the science of relating stellar oscillations to their structural origin, thrived. Furthermore, due to the excellent precision of the data at hand, it was made possible to detect extremely faint features and take advantage of these to constrain the stellar structure. Acoustic glitches are such features and present themselves as an oscillating signal in the observed frequencies \citep[e.g.][]{2007MNRAS.375..861H}. Although it was predicted that such signatures could be detected for solar-like stars other than the Sun \citep{1998IAUS..185..317M}, due to lack of precision in the data, they had only been observed in the solar case \citep[with the first mention of such signatures as early as almost four decades ago][]{1986Ap&SS.126..335H,1988IAUS..123..151V,1990LNP...367..283G}. As these glitches are caused by sharp variations in the stellar structure, they hold valuable and localised information. For example, the glitch caused by the second ionisation zone of helium carries information about the surface helium content. Indeed, \cite{2004MNRAS.350..277B} demonstrated a positive correlation between the strength of the helium glitch signal and the envelope helium abundance in the solar case. Therefore, it allows us to lift the degeneracy between the stellar mass and helium abundance of low mass stars uncovered by \cite{2014A&A...569A..21L} that greatly reduces the precision of stellar models. As a consequence, many studies interested themselves in such glitches, for example \cite{2021A&A...655A..85H,2022A&A...663A..60H} theoretically related the properties of the ionisation region (hydrogen, first and second partial helium) to the glitch's signature in a model independent fashion. Other approaches \citep[such as][]{2002ASPC..274...77M,2004MNRAS.350..277B,2014ApJ...782...18M,2014ApJ...790..138V,2017ApJ...837...47V,2022MNRAS.515.1492V,2019A&A...622A..98F} focused on providing means to retrieve the helium glitch signature and building accurate models reproducing this signature. 

With upcoming missions such as PLATO \citep{2014ExA....38..249R} striving to accurately characterise low-mass stars, the automated and swift retrieval of the glitch parameters is an essential milestone. Nevertheless, this task is arduous and may require advanced and rather slow techniques. The cause is the non-linear nature of the function that is adjusted to the observed frequencies \citep[i.e.][]{2007MNRAS.375..861H,2014ApJ...790..138V}. To address these issues and provide a robust adjustment of the glitch, in order to accurately provide a constraint on the surface helium content, \cite{2019A&A...622A..98F} provided a linearised approach, which has the advantage of being extremely fast (only a fraction of a second per star) and provides constraints to the stellar structure which show very little correlation, the \who\footnote{\url{https://github.com/Yuglut/WhoSGlAd-python}} (\whoac) method. However, their technique requires an estimate for the acoustic depth at which the helium glitch occurs. This leads to several difficulties. Firstly, while physically motivated by previous studies, the exact glitch formulation used in \cite{2019A&A...622A..98F} (and other similar ones) is determined empirically and does not directly relate to the properties of the glitch it aims at representing. Secondly, the actual definition of the acoustic depth of the helium glitch is somewhat arbitrary. For example, as illustrated in \cite[Figs.~1 and 3, respectively]{2014ApJ...790..138V,2021A&A...655A..85H}, the depression in the first adiabatic index $\Gamma_1$, causing the helium glitch signature, is actually the composite effect of the hydrogen ionisation and helium first and second ionisations. The first adiabatic index is defined as
\begin{equation}
    \Gamma_1~\equiv~\left.\frac{d \ln P}{d \ln \rho}\right\vert_S,
\end{equation}
with $P$ the pressure, $\rho$ the density, and $S$ the entropy. Consequently, the $\Gamma_1$ depression presents a broad feature and defining its exact position is somewhat arbitrary, which may impact the retrieved final glitch properties. Both \cite{2014ApJ...790..138V,2019A&A...622A..98F} use the depth of the peak between the first and second helium ionisation zones. Additionally, the \who approach defined in \cite{2019A&A...622A..98F} relies on stellar models to provide a value for this depth in order to preserve the linearity of the method. This hampers the ability of the method to be automated as preliminary stellar models have to be built before the helium glitch signature can be adjusted. Another consequence is that the adjusted glitch signature becomes somewhat model dependent. This is again something one aims to avoid as stellar models themselves present a large number of uncertainties (e.g. reference solar mixture, mixing prescription, opacities,...). Following on from \cite{2019A&A...622A..98F}, we aim in this paper at studying the evolution of the helium acoustic depth across the seismic HR diagram (defined using \who indicators) and the impact of different approaches for its determination on the glitch amplitude and its use as a seismic indicator. This is motivated by the fact that the second ionisation of helium is mostly determined by the local temperature and density, as one would expect from Saha's relation \citep{doi:10.1080/14786441008636148}. We then provide a prescription to automatically estimate the helium glitch acoustic depth and assess the impact of this prescription on the inferred helium abundance, crucial to the accurate determination of stellar parameters. This new prescription renders \cite{2019A&A...622A..98F}'s \who method completely automatic and makes it a swift candidate for the survey of the glitches of large samples of solar-like stars as is expected of the PLATO mission \citep{2014ExA....38..249R}.

This paper is organised as follows: In Sect.~\ref{Sec:Mot} we detail the motivation behind our approach and recall the basic principle behind the \who method. We pursue with our theoretical results on a grid of models in Sect.~\ref{Sec:Th}. We then apply the developed approach to a set of four observed targets in Sect.~\ref{Sec:App} and characterise both its accuracy and efficiency. Finally, we conclude our paper in Sect.~\ref{Sec:Con}.

\section{Motivation}\label{Sec:Mot}
\subsection{\who fitting reminder}
The peculiarity of the \who method lies in the definition of the fitting function. An orthonormal basis of functions is built over the vector space of frequencies. These functions are separated into a smooth -- slowly varying trend as a function of frequency, as per the asymptotic theory \citep[e.g.][]{1980ApJS...43..469T} -- and a glitch contribution, which are independent of one another. Indeed, by construction, the basis functions used to represent the frequencies are totally orthonormal to those of the smooth part.\footnote{We note that it is possible, from a physical point of view, that the glitch functions may contain some information related to the smooth contribution and conversely, although they have been selected to minimise such effect. We merely mean that, by construction, the functions used to represent both contributions are mathematically independent of one another. It is clear that the oscillation frequencies of a star are not physically independent. However, if their \emph{measurements} are treated as independent probability variables, then our orthogonalisation ensures that the measured seismic indicators associated with the smooth component are statistically independent from those associated to the glitch(es).} This has the advantage of rendering the indicators defined over the glitch contribution completely independent of their smooth counterparts. The general representation of a fitted frequency of radial order $n$ and spherical degree $l$ is the following
\begin{equation}
    \nu_{n,l,\textrm{fit}} = \sum\limits_k a_{k} q_k\left(n\right),
\end{equation}
with $a_{k}$ the projected reference frequency over the basis function of index $k$, $q_k$, evaluated at $n$. These orthonomal functions are obtained by applying the Gram-Schmidt algorithm to the polynomials of increasing degrees $n^k$ for the smooth part and a parametrised  oscillating component for the glitch \citep[see][and Eq.~\ref{Eq:pHe} below]{2019A&A...622A..98F} The projection is done for each spherical degree according to the scalar product defined in \cite{2019A&A...622A..98F}. The specifity of this definition of the scalar product is that it accounts for the observational uncertainties on the oscillation frequencies.

Because of the orthornormalisation, the $a_k$ coefficients are completely independent of one another and of unit uncertainty. By consequence, combining them appropriately allows us to construct indicators that are as little correlated as possible. Central to the present discussion is the helium glitch amplitude indicator, $A_{\textrm{He}}$, which has the advantage of being completely independent of the indicators defined over the smooth part of the oscillation spectrum. In the present paper, we propose a revised definition of the amplitude defined in \cite{2019A&A...622A..98F} that scales with the uncertainties on the observed frequencies
\begin{equation}
    \mathcal{A}_{\textrm{He}} = \frac{A_{\textrm{He}}}{\sqrt{\sum\limits_{i=1}^N1/\sigma^2_i}},
\end{equation}
with $A_{\textrm{He}}$ the helium glitch amplitude as defined in \cite{2019A&A...622A..98F} and $\sigma_i$ the uncertainty on the i-th of the N observed frequencies.
Nevertheless, to retrieve the helium glitch signature, fitting methods need to determine the acoustic depth at which the glitch occurs \citep[see for example][]{2002ASPC..274...77M,2004MNRAS.350..277B,2007MNRAS.375..861H,2014ApJ...790..138V}. It is defined as,
\begin{equation}\label{Eq:TauMod}
    \tau_{\textrm{He}} = \int\limits_{r_{\textrm{He}}}^{R_*} \frac{dr}{c(r)},
\end{equation}
where $r$ is the radius of the considered layer of the star, $R_*$ the radius of the stellar surface, $r_{\textrm{He}}$ the radius of the helium glitch, and $c\left(r\right)$ the local sound speed. The exact value of $r_{\textrm{He}}$ is somewhat arbitrary as the depression in $\Gamma_1$ corresponds to the contribution of the hydrogen, first and second partial helium ionisation zones \citep[See for example Figs.~1 and 3 in][]{2014ApJ...790..138V,2021A&A...655A..85H}. Determining its value therefore constitutes an uncertainty of glitch fitting approaches.
The \who method is no exception as the glitch orthonormal basis elements are function of this acoustic depth. For clarity, we recall the basis elements used to describe the helium glitch (before orthonormalisation)
\begin{equation}
    p_{\textrm{He},k,j}(\widetilde{n})~=~f_j\left( 4\pi T_\textrm{He}\widetilde{n}\right) \widetilde{n}^{~-k},
    \label{Eq:pHe}
\end{equation}
with $k~=~(4,5)$, $j~=~(1,2)$, $f_1\left(\bullet\right)~=~\sin\left(\bullet\right)$, $f_2\left(\bullet\right)~=~\cos\left(\bullet\right)$, $\widetilde{n}~=~n+l/2$, and 
\begin{equation}\label{Eq:TauAdim}
    T_{\textrm{He}} = \tau_{\textrm{He}} \Delta,
\end{equation}
the dimensionless acoustic depth of the helium glitch where $\Delta$ is the large frequency separation as defined in \who. This redefinition of the acoustic depth prevents the frequency function to be fitted from being implicit in frequency and further reduces non-linearities. This $T_{\textrm{He}}$ parameter remains the only non-linear parameter present in the \who formulation. \cite{2019A&A...622A..98F,2020A&A...644A..37F} resolve this issue by keeping it fixed to a model value obtained from a partial modelling. While their method has proven efficient and accurate for the 16CygA and B system, the need for a $T_{\textrm{He}}$ value, which is model dependent, prevents their approach from being fully automated. This is the issue we aim to address in the present publication.

\subsection{Some intuition}
The objective of the present paper is to relate the dimensionless helium glitch acoustic depth to observables that are easy to obtain. To do so, we may build some intuition from Saha's relation \citep{doi:10.1080/14786441008636148}
\begin{equation}\label{Eq:Saha}
    \frac{He^{++} n_e}{He^+}~=~\frac{g}{h^3}\left(2\pi m_e k_B T\right)^\frac{3}{2} e^{-\frac{\chi}{k_B T}},
\end{equation}
where $g$ is the ratio of the statistical weights between the first and second ionisation states of helium, $h$ Planck's constant, $k_B$ Boltzmann's constant, $m_e$ the electron mass, and $\chi$ the helium second ionisation energy. This equation allows us to compute the number of fully ionised helium atoms $He^{++}$, that of partially ionised helium atoms $He^{+}$ and the number of free electrons $n_e$, which actually conceals a dependency in density. While we assume here that helium is the only species to be partially ionised -- hydrogen is fully ionised and metals are not --, Eq.~\ref{Eq:Saha} shows that both the temperature and density at the layer impact the relative number of helium atoms in their first and second states of ionisation. Therefore, we expect the acoustic depth of the glitch to be mostly determined by the profile in the stellar interior of these two quantities. The exact nature of this profile should vary with stellar parameters. Therefore, the first natural step of this work is to study the evolution of the dimensionless acoustic depth of the helium glitch (Eq.~\ref{Eq:TauAdim}) across the HR diagram, where the effective temperature directly intervenes. Rather than using the classical $T_{\textrm{eff}}$ - $L$ diagram, we take advantage of the precise seismic data and build a seismic HR diagram as in \cite{1988IAUS..123..295C} but using the $\Delta_0$ and $\hat{r}_{02}$ seismic indicators \citep[as defined in][]{2019A&A...622A..98F}.  These correspond to the large separation of radial modes and the average small separation ratio between radial and quadrupolar modes.

\section{Theoretical study of the helium acoustic depth over a grid of models and linear estimation}\label{Sec:Th}

\subsection{Evolution across the seismic HR diagram}
To motivate the use of a linear estimation of $T_{\textrm{He}}$, we show its evolution on the main sequence along a grid of models with masses ranging from $0.9~M_{\odot}$ to $1.18~M_{\odot}$. We selected this range to focus on models that do not have a convective core on the main sequence, which could introduce higher order contributions in the acoustic depth evolution. (We indeed observed for higher masses that the helium acoustic depth varies in a strongly non-linear fashion as a function of $\Delta$ and $\hat{r}_{02}$, as opposed to the lower masses.) All the models have been built using the CLES stellar evolution code \citep{2008Ap&SS.316...83S} as described in \cite{2019A&A...622A..98F}.

\begin{figure}
\includegraphics[width=\linewidth]{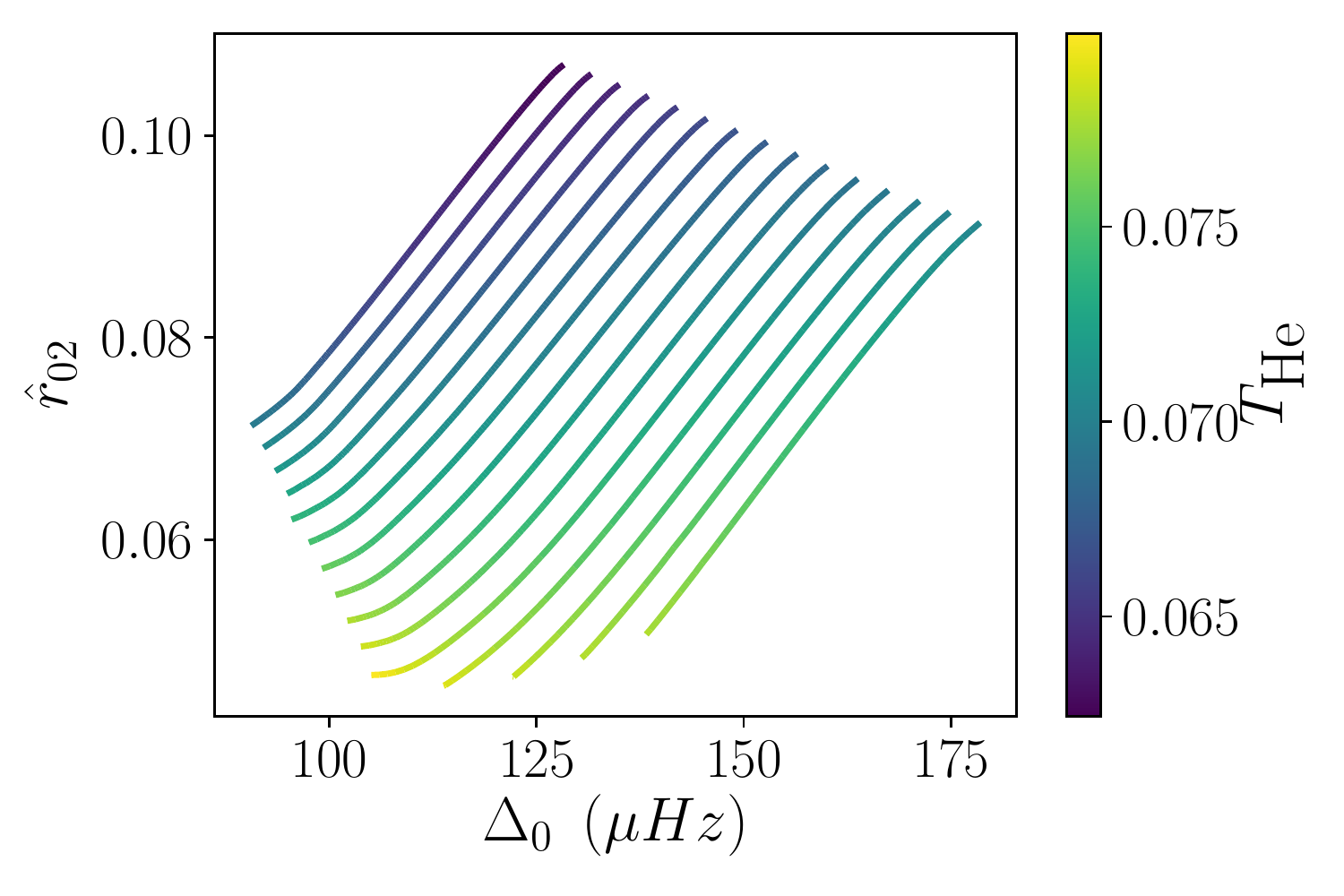}
\caption{Evolution during the main sequence of the dimensionless helium glitch acoustic depth along a grid of models. The models have an initial composition of $X_0=0.68$ and $Z_0=0.024$ and masses ranging from $0.90M_{\odot}$ to $1.18M_{\odot}$ ($0.02M_{\odot}$ step, right to left). The colour gradient represents the value of $T_{\textrm{He}}$.}\label{Fig:TauEvol}
\end{figure}

Figure \ref{Fig:TauEvol} shows the evolution of the dimensionless acoustic depth, as a colour gradient, with $\Delta_0$ and $\hat{r}_{02}$ for a composition of $X_0=0.68$ and $Z_0=0.024$, which is typical of solar-like stars. In this figure, each track corresponds to an evolutionary track on the main sequence (starting at the ZAMS on the top right and finishing at the TAMS on the bottom left) for a given mass, increasing from right to left. We first observe that its evolution along the grid seems both monotonic and linear, with increasing values from top to bottom. Therefore, we adjust a linear relation in $\Delta_0$ and $\hat{r}_{02}$ of the form
\begin{equation}\label{Eq:TauFit}
T_{\textrm{He,lin}} \simeq a \Delta_0 + b \hat{r}_{02} + c,
\end{equation}
where $a$, $b$, and $c$ are the coefficients to be adjusted.

\begin{figure}
\includegraphics[width=\linewidth]{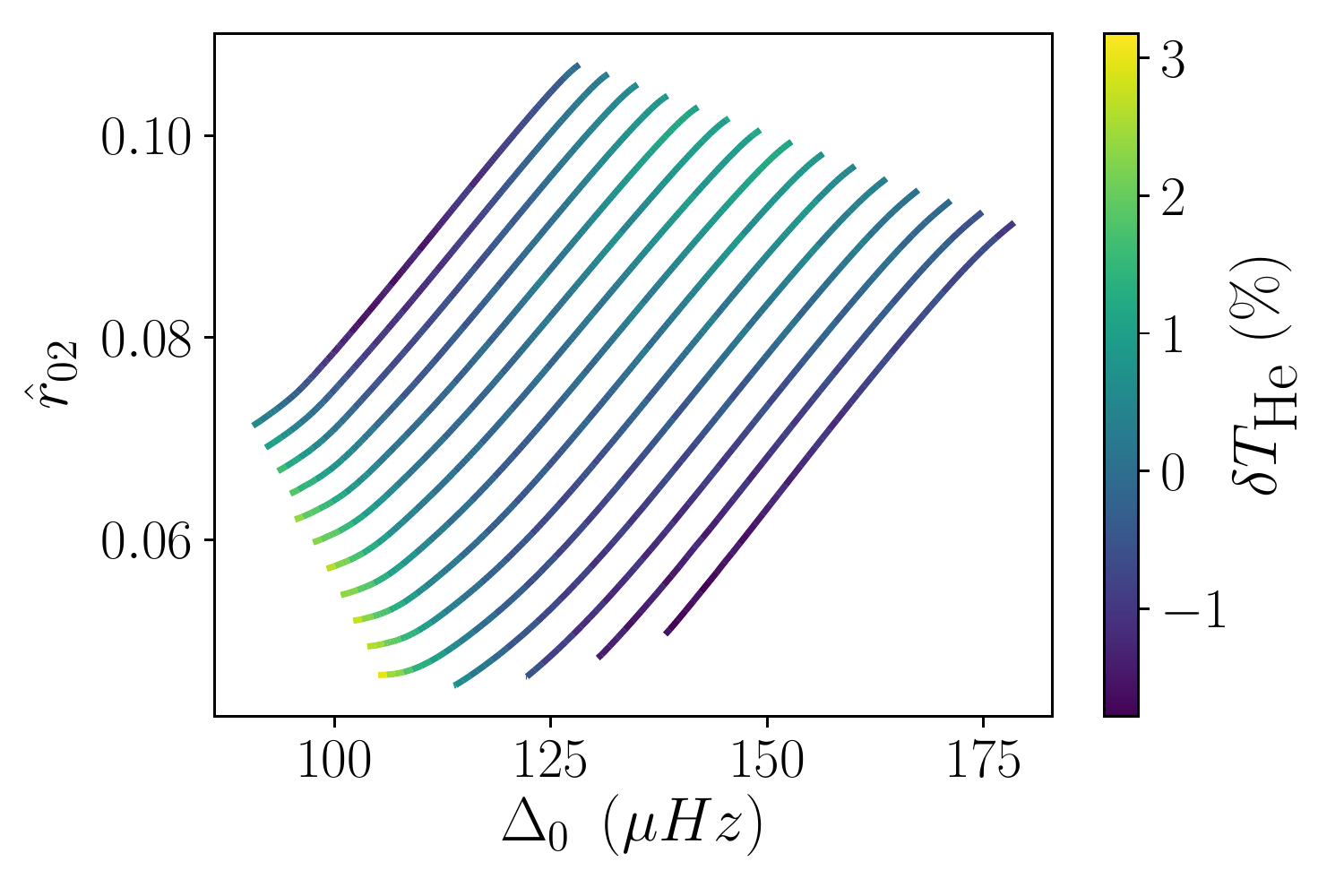}
\caption{Evolution during the main sequence of the difference between the fitted dimensionless helium glitch acoustic depth along over the same grid of models as in Fig.~\ref{Fig:TauEvol}. The colour gradient represents the relative difference in $T_{\textrm{He}}$ between the model and linearly adjusted values, expressed in $\%$.}\label{Fig:TauDiff}
\end{figure}

To validate our adjustment, we now compute the reduced differences in dimensionless acoustic depth between the fitted and exact (with the elected location of the helium glitch) values over the same grid. These are expressed as
\begin{equation}\label{Eq:TauDiff}
    \delta T_{\textrm{He}} = \frac{T_{\textrm{He,model}}-T_{\textrm{He,lin}}}{T_{\textrm{He,model}}},
\end{equation}
with the `model' and `lin' subscripts representing the model value obtained by integration (Eqs.~\ref{Eq:TauMod} and \ref{Eq:TauAdim}) and the one obtained by the adjusted linear relation (Eq.~\ref{Eq:TauFit}), respectively. This is represented as a colour gradient in Fig.~\ref{Fig:TauDiff} where the difference is expressed in percentage. We observe that the relation fares rather well with a maximum discrepancy of at most $3~\%$ (which corresponds to an absolute error of $T_{\textrm{He,model}}-T_{\textrm{He,lin}} \simeq 0.0025$, about a tenth of the $\sim 0.02$ span of the grid). For the sake of comparison, \cite{2019A&A...622A..98F} had stated that, in 16CygA's case, a change in the dimensionless acoustic depth of $10~\%$ had no significant impact on the measured glitch amplitude, meaning that the inferred helium abundance would remain untouched. This however has to be regarded with caution as it corresponds only to a specific star and the physical conditions can substantially vary from star to star. We also note in Fig.~\ref{Fig:TauDiff} that the difference does not vary monotonically along the grid, showing an hourglass shape with a zero crossing close to its center and maximum differences at the zero-age main sequence and terminal-age main sequence. This is probably a calibration effect and suggests that a higher order relation could improve the agreement. However, given the already satisfactory agreement, such more complex relations are unnecessary. Furthermore, we will later touch upon means to further improve our results (see Sect.~\ref{Sec:App}).

\subsection{Impact of the composition}
\begin{figure}
\includegraphics[width=\linewidth]{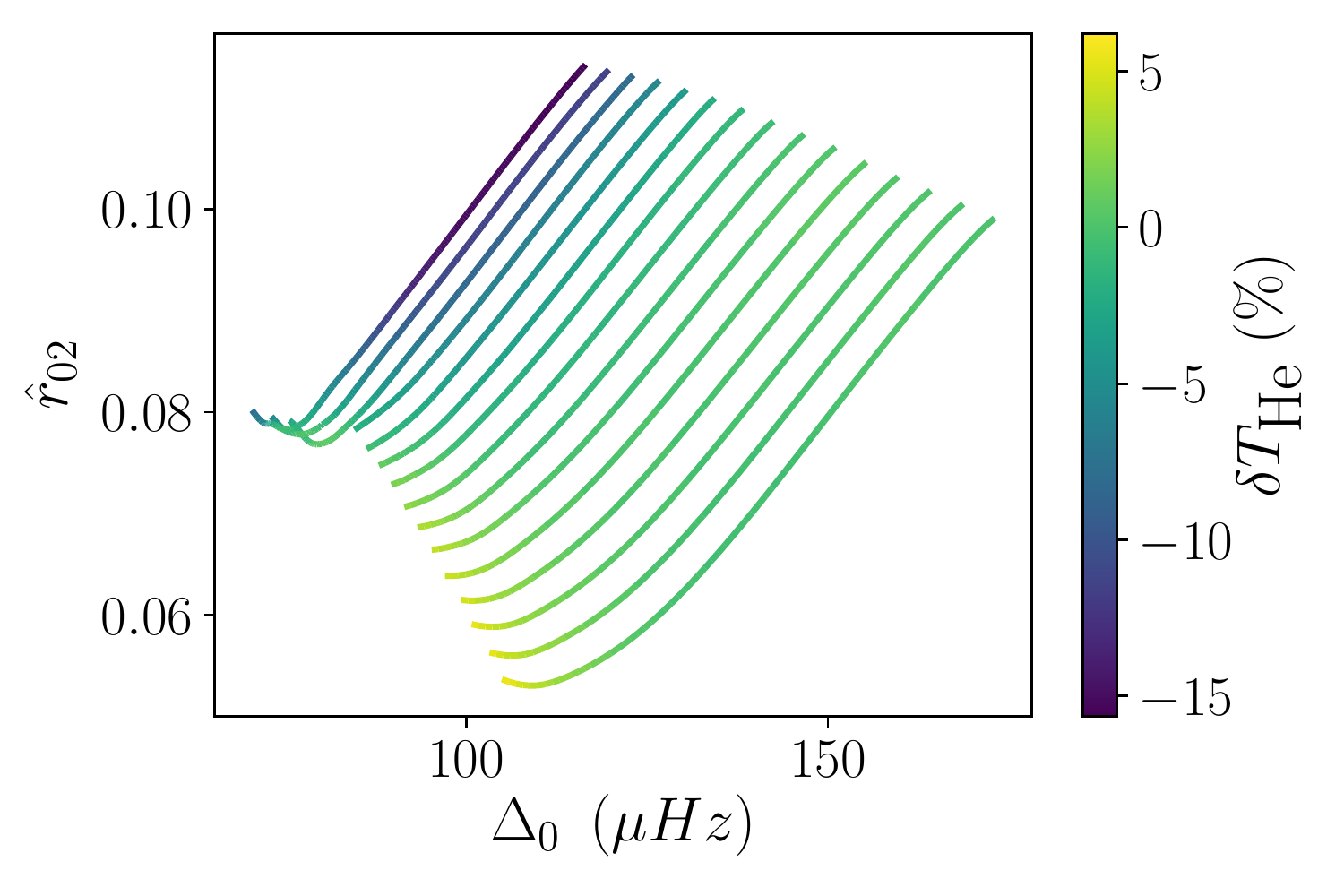}
\caption{Same as Fig.~\ref{Fig:TauDiff} for an initial composition of $X_0=0.68$ and $Z_0=0.012$.}\label{Fig:TauDiffX68Z12}
\end{figure}

\begin{figure}
\includegraphics[width=\linewidth]{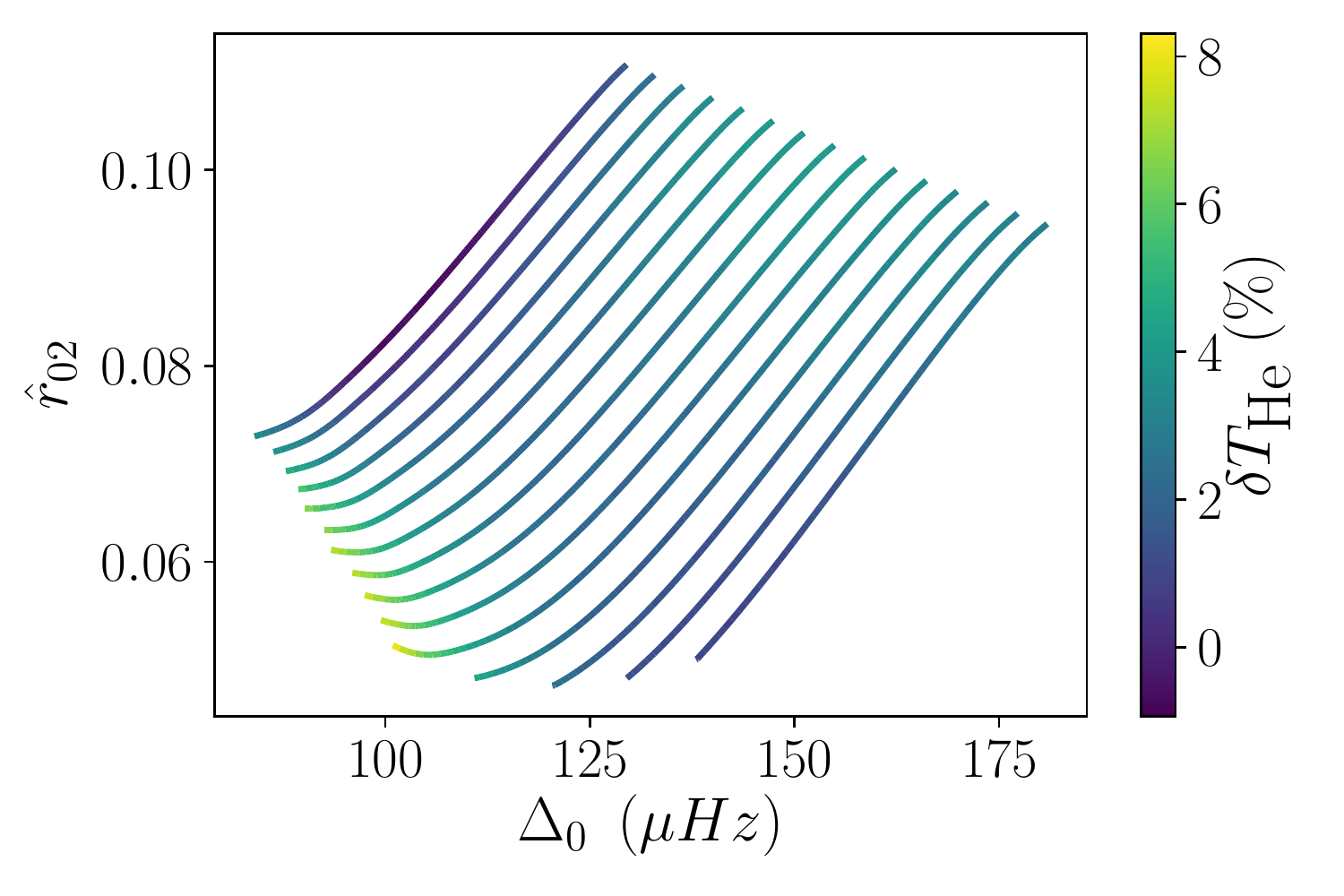}
\caption{Same as Fig.~\ref{Fig:TauDiff} for an initial composition of $X_0=0.74$ and $Z_0=0.012$.}\label{Fig:TauDiffX74Z12}
\end{figure}

\begin{figure}
\includegraphics[width=\linewidth]{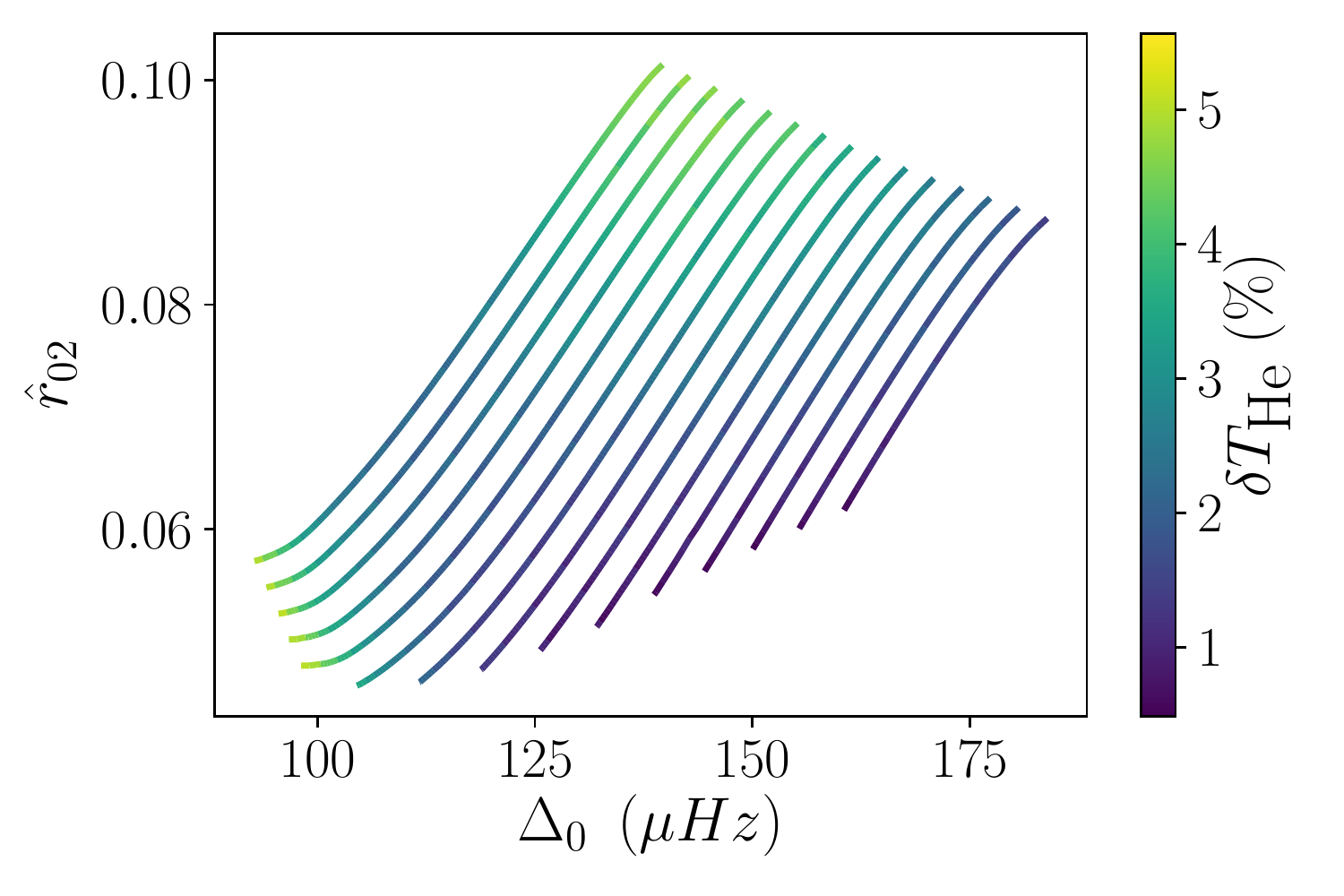}
\caption{Same as Fig.~\ref{Fig:TauDiff} for an initial composition of $X_0=0.74$ and $Z_0=0.024$.}\label{Fig:TauDiffX74Z24}
\end{figure}

As one would expect, the stellar composition may vary between stars and we expect that it will impact the measured glitch acoustic depth. To test this hypothesis, we compute the relative difference in acoustic depth (Eq.~\ref{Eq:TauDiff}) over grids of models with different chemical compositions. We consider here the composition pairs $\left(X_0,Z_0\right)=(0.68,0.012)$, $\left(0.74,0.012\right)$, and $\left(0.74,0.024\right)$ which should display large enough variations to observe an impact on the estimated helium glitch acoustic depth \citep[e.g. typical ranges of values such as the one considered by][ Fig.~4, encompass the one considered here]{2021MNRAS.500...54N}. We illustrate these results in Figs.~\ref{Fig:TauDiffX68Z12} to \ref{Fig:TauDiffX74Z24} where we use the values of the coefficients fitted to our reference case ($X_0=0.68$ and $Z_0=0.024$, Figs.~\ref{Fig:TauEvol} and \ref{Fig:TauDiff}).

\begin{figure}
\includegraphics[width=\linewidth]{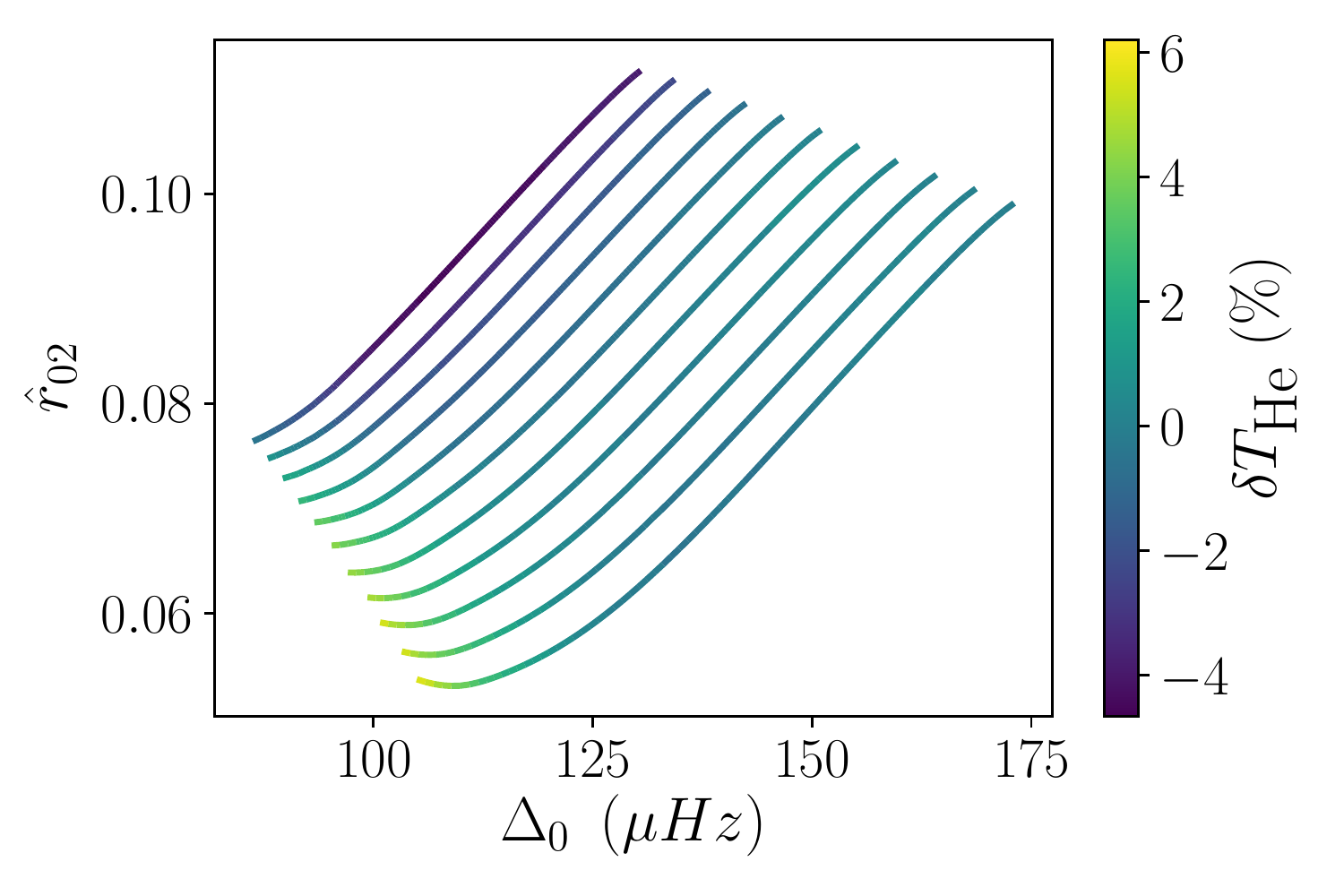}
\caption{Same as Fig.~\ref{Fig:TauDiffX68Z12} with a reduced mass range ($M\in\left[0.90M_{\odot},1.12M_{\odot}\right]$) to include only models without a convective core on the main sequence.}\label{Fig:TauDiffX68Z12-NoConv}
\end{figure}

We observe that the differences are greater than for our reference case reaching down to $-15~\%$ in the worst case ($X_0=0.68$, $Z_0=0.012$). It is therefore necessary to further assess the impact of the stellar composition on the estimated acoustic depth. To do so, we adjust Eq.~\ref{Eq:TauFit} for each of these grids. The adjusted parameters for all the considered compositions are summarised in Table~\ref{Tab:par}, where the reference case is displayed in boldface for clarity. We note in this table that the fitted coefficients may dramatically change, in some cases almost tripling in value. However, considering values typical of the middle of our grid, this change in fitted coefficients with composition, only leads to an approximate difference of $\sim~0.3\%$ in $T_{\textrm{He}}$ with respect to our reference case. We also add that in the worst case ($X_0=0.68$, $Z_0=0.012$, Fig.~\ref{Fig:TauDiffX68Z12}), only the models with a mass greater than $1.12~M_{\odot}$ present relative differences as high as $\sim 15\%$. These actually correspond to models that preserve a convective core from their pre-main sequence phase, while stellar models of lower mass do not. This introduces non-linearities which may explain the large differences we observe in Fig.~\ref{Fig:TauDiffX68Z12}. Indeed, the large discrepancies only appear for these masses while the other tracks are confined in a smaller range around zero. Removing tracks with a convective core from Fig.~\ref{Fig:TauDiffX68Z12} indeed reduces the discrepancies as displayed in Fig.~\ref{Fig:TauDiffX68Z12-NoConv}, where the differences now span the $-4\%$ to $6\%$ range. Consequently, all the values computed for models without convective cores are within $10\%$ of the actual value. This demonstrates that, on the main sequence and for models without a convective core, the linear estimate should provide results of reasonable quality.

\begin{table}
\caption{Fitted coefficients for Eq. \ref{Eq:TauFit} at different compositions. The reference case is shown in boldface. Note, for some compositions (e.g. 0.74, 0.024), our criterion to stop the evolutionary tracks ($10~\textrm{Gyr}$) lead to an early termination that may slightly bias the values.}
\centering
\begin{tabular}{cc|ccc}\label{Tab:par}
X & Z & a ($Ms$) & b & c \\
\hline
$0.68$ & $0.012$ & $1.6E-04$ & $-4.3E-01$ & $8.4E-02$ \\
$\mathbf{0.68}$ & $\mathbf{0.024}$ & $\mathbf{8.6E-05}$ & $\mathbf{-2.8E-01}$ & $\mathbf{8.1E-02}$ \\
$0.74$ & $0.012$ & $1.0E-04$ & $-3.1E-01$ & $8.4E-02$ \\
$0.74$ & $0.024$ & $6.2E-05$ & $-2.3E-01$ & $8.3E-02$ 
\end{tabular}
\end{table}

\section{Performance of different estimates with observed targets and impact on the inferred helium abundance}\label{Sec:App}
In the present section, we compare the results of three different approaches to estimate the helium glitch acoustic depth and assess their ability to reliably estimate the helium glitch amplitude, to in turn model the helium abundance. These three approaches are the linear formulation in $\Delta$ and $\hat{r}_{02}$ (Eq.~\ref{Eq:TauFit}), an optimised value via Brent's optimisation algorithm (minimising the differences between reference and fitted frequencies) and the value obtained by partial modelling of the target and computation of the integrated form, Eq.~\ref{Eq:TauMod} (as formerly used with \who) -- dubbed `old' in the present paper. To do so, we consider four Kepler targets: 16CygA, 16CygB, KIC8006161, and KIC8394589. These are all solar-like stars that are within the Kepler Legacy sample \citep{2017ApJ...835..172L} which constitute the most precise seismic data to this day. Except for 16CygA and B for which we use the revised frequencies of \cite{2015MNRAS.446.2959D}, we use the frequencies stated in \cite{2017ApJ...835..172L}. From these frequencies, we ignore the ones with uncertainties greater than $1~\mu Hz$ as they are the most uncertain and may destabilise the search for models representative of \who indicators.

\subsection{Accuracy of the linear estimate}
\begin{figure}
    \centering
    \includegraphics[width=\linewidth]{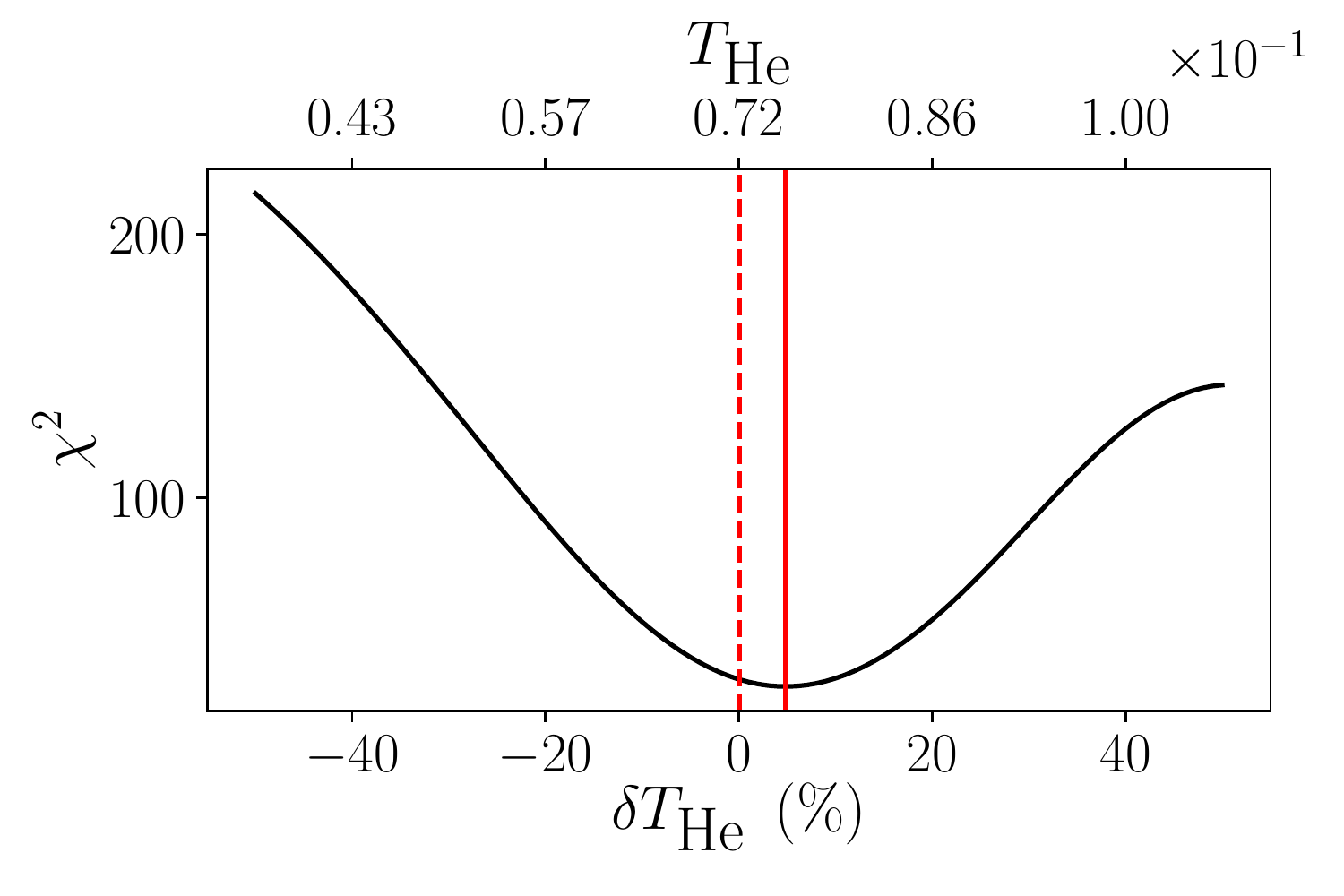}
    \caption{Evolution of the agreement between fitted and observed frequencies of 16CygB, expressed as a $\chi^2$ value, as a function of the helium glitch acoustic depth, expressed as the relative difference with respect to the linear estimate (in $\%$). The $\chi^2$ minimum, obtained via Brent's minimisation procedure, is represented by the red vertical line. We also show the linear estimate as a vertical dashed line and the value obtained with the `old' approach as a dotted one (due to its proximity with the continuous one, it is barely visible).}
    \label{Fig:16CygB-chi}
\end{figure}

Before assessing whether the linear estimate of the helium acoustic depth allows us to retrieve an accurate helium abundance, we assess the accuracy of its value. To do so, we plot the evolution of the $\chi^2$ function -- evaluating the agreement between observed and fitted frequencies -- with the value of the dimensionless acoustic depth. This function is defined as
\begin{equation}
    \chi^2 = \sum\limits^N_{i=1} \frac{\left(\nu_{i,\textrm{obs}} - \nu_{i,\textrm{fit}}\right)^2}{\sigma_i^2},
\end{equation}
where $\nu_i$ is the i-th of the N frequencies, the `obs' and `fit' subscript corresponding to observed and \who fitted values, respectively, and $\sigma_i$ the uncertainty of the i-th frequency. We show the two characteristic cases of 16CygB and KIC8394589 in Figs.~\ref{Fig:16CygB-chi} and \ref{Fig:8394589-chi}, which constitute two extremes (the plots for the two remaining stars are given in Appendix~\ref{Sec:app-chi}). In both figures, the horizontal axis corresponds to the value of the dimensionless acoustic depth, either as a relative difference from the linear estimate expressed as a percentage at the bottom, or as the exact value shown at the top. Because the bottom axis corresponds to a relative difference with respect to the linear estimate derived in the present paper, a zero value corresponds to this estimate, shown as the red dashed line for clarity. In both figures, we consider values in the range $\left[0.5 ~ T_{\textrm{He,lin}},1.5 ~ T_{\textrm{He,lin}} \right]$, with the linear estimate denoted as $T_{\textrm{He,lin}}$.

Focusing first on Fig.~\ref{Fig:16CygB-chi}, we observe that the $\chi^2$ is well behaved and presents a clear minimum. In addition, we note that this minimum is approximately $5\%$ from the linear estimate. This means that the linear estimate provides a good guess of the optimal value. Figure~\ref{Fig:8394589-chi} depicts a slightly more complicated picture for two reasons. First, the $\chi^2$ landscape produces a less clear minimum, due to a flat region at the bottom of the $\chi^2$ depression. Second, we observe now that this minimum lies at almost $1.4$ times the linearly computed value. This could have a significant impact on the inferred helium abundance. This will be assessed in Sect.~\ref{Sec:HeEst}.

Leaning on these considerations, we improve the robustness of our approach by optimising over the dimensionless acoustic depth value. That is, we aim at improving the agreement between the \who adjusted frequencies and the reference ones and therefore at decreasing the $\chi^2$ value by iterating over the $T_{\textrm{He}}$ parameter. As we expect the $\chi^2$ landscape to locally resemble a parabola close to its minimum, as shown in Figs.~\ref{Fig:16CygB-chi} and \ref{Fig:8394589-chi}, we use Brent's minimisation algorithm to optimise over $T_{\textrm{He}}$, taking as a starting point the linear estimate developed in the present paper. Due to the swiftness of the method, this can be done within a fraction of a second. To illustrate the results of the minimisation, we show the optimised values as a red vertical line in Figs.~\ref{Fig:16CygB-chi} and \ref{Fig:8394589-chi}. This is also a clear illustration that Brent's method reached the expected minimum.

\begin{figure}
    \centering
    \includegraphics[width=\linewidth]{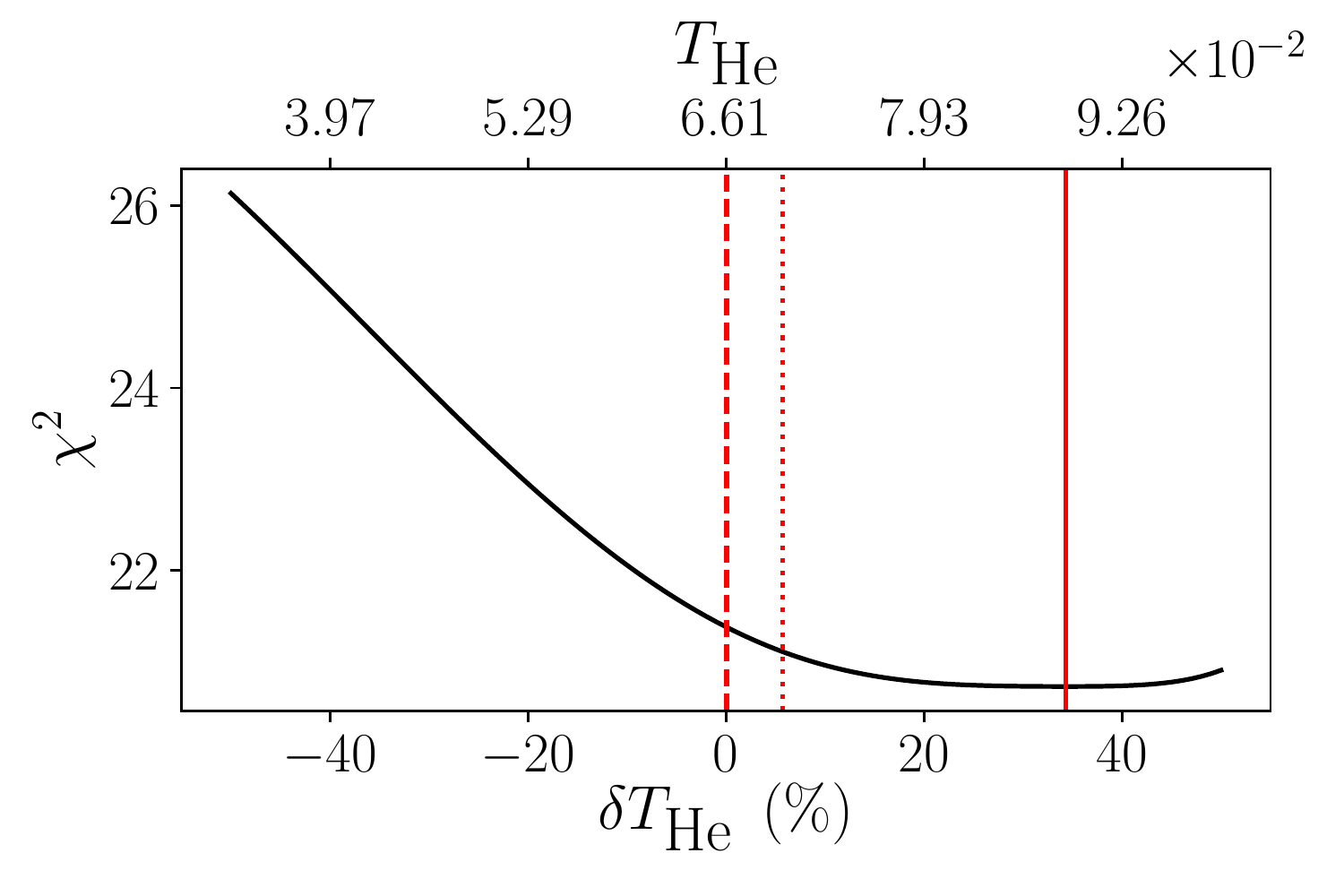}
    \caption{Same as Fig.~\ref{Fig:16CygB-chi} in KIC8394589's case.}
    \label{Fig:8394589-chi}
\end{figure}

\subsection{Impact of the helium acoustic depth on the measured glitch amplitude}\label{sec:AHe}
One of the most crucial end products of helium glitch fitting is the ability to precisely constrain the helium abundance of solar-like stars. As we have provided several estimates for $T_{\textrm{He}}$, which may in turn impact the helium glitch amplitude and, therefore the inferred helium abundance, it becomes necessary to compare these approaches and assess their accuracy when modelling the helium abundance. We first assess the impact of the estimated value of $T_{\textrm{He}}$ on the measured glitch amplitude, $\mathcal{A}_{\textrm{He}}$. Using the same set of benchmark stars as previously, we compute the evolution of the helium glitch amplitude over the same range of acoustic depth values as in Figs.~\ref{Fig:16CygB-chi} and \ref{Fig:8394589-chi}. The results are presented in Figs.~\ref{Fig:16CygB-AHe} and \ref{Fig:8394589-AHe} (additional plots are presented in Appendix~\ref{Sec:app-chi}). In these figures, the value corresponding to the optimum of $\chi^2$ -- as retrieved with Brent's optimisation procedure --, is shown by the continuous red vertical line. The meaning of the axes and vertical lines are the same as in Figs.~\ref{Fig:16CygB-chi} and \ref{Fig:8394589-chi}, with the dotted line corresponding to the `old' approach and the dashed line to the linear estimate presented in this paper. Individual estimates of the helium glitch acoustic depth are provided in Table~\ref{Tab:THeComp}. Additionally, for better visualisation purposes, we show a departure of one $\sigma\left(\mathcal{A}_{\textrm{He}}\right)$ with respect to the optimised value as the horizontal blue line. Observing these figures, it is striking that all the $T_{\textrm{He}}$ estimates provide a measurement of the glitch amplitude within one $\sigma$ of the optimised value, which we would consider to be the better one. Only a large change in $T_{\textrm{He}}$ would lead to a significant change in $\mathcal{A}_{\textrm{He}}$. Consequently, these results demonstrate the relative insensitivity of our approach to the exact value of this somewhat arbitrary parameter -- we recall that its exact definition is mostly a matter of convention as the feature from which the helium glitch originates is rather broad. From these considerations, we expect that a simple approach is best for the swiftness and automation of \who as it should not significantly impact the inferred helium abundance. Therefore, we recommend to use the value obtained by Brent's optimisation scheme as it is easy to implement, fast in execution, robust, and relatively model independent.

\begin{figure}
    \centering
    \includegraphics[width=\linewidth]{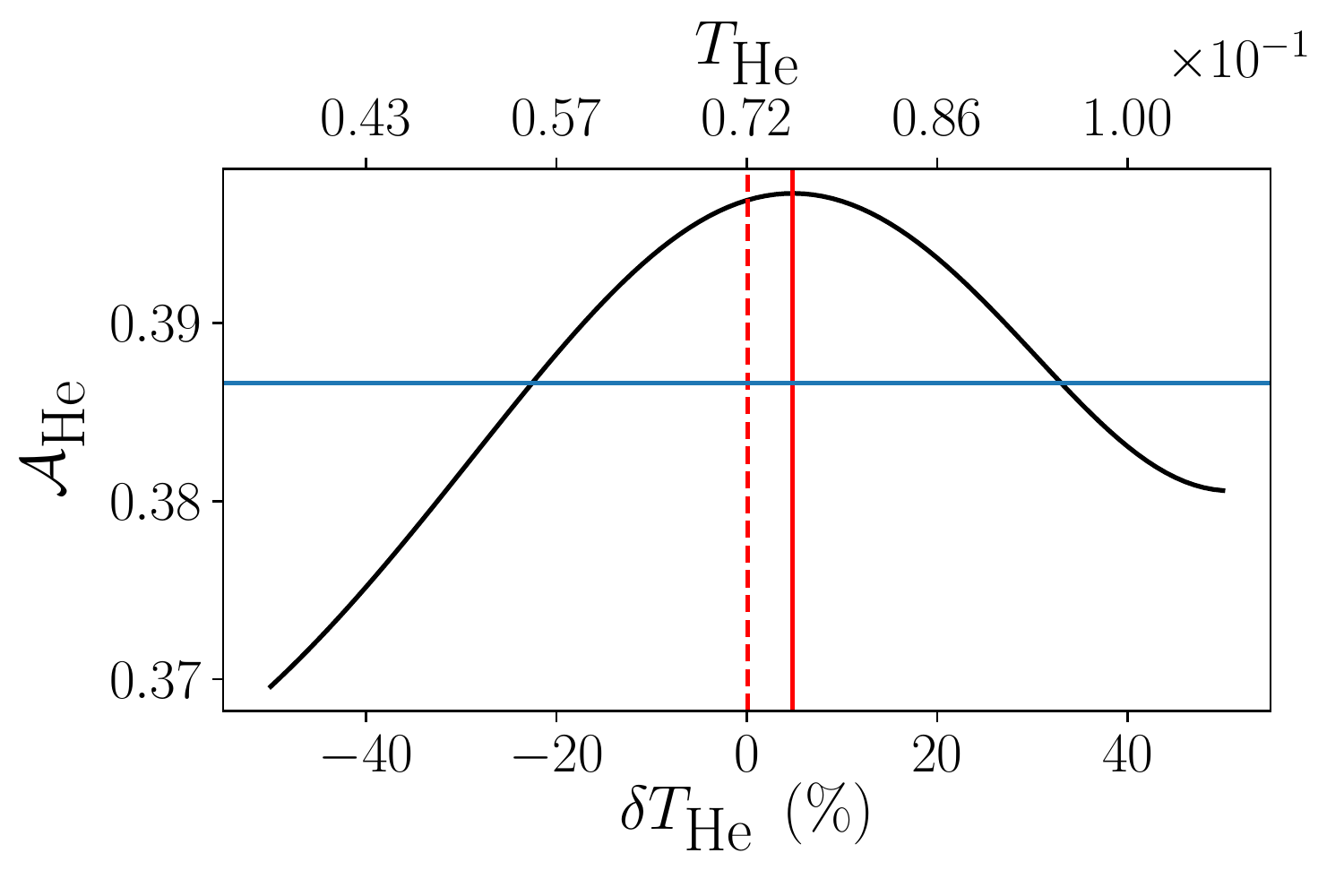}
    \caption{Evolution of the helium glitch amplitude of 16CygB with the variation of the helium glitch acoustic depth with respect to the linear estimate (Eq.~\ref{Eq:TauFit}). This variation is expressed as a relative difference in percent. The $\chi^2$ minimum, obtained via Brent's minimisation procedure, is represented by the red vertical line. We also show the linear estimate as a vertical dashed line and the value obtained with the `old' approach as a dotted one (due to its proximity with the continuous one, it is barely visible). The blue horizontal line corresponds to a $1~\sigma$ variation in amplitude from the optimal value.}
    \label{Fig:16CygB-AHe}
\end{figure}

\begin{figure}
    \centering
    \includegraphics[width=\linewidth]{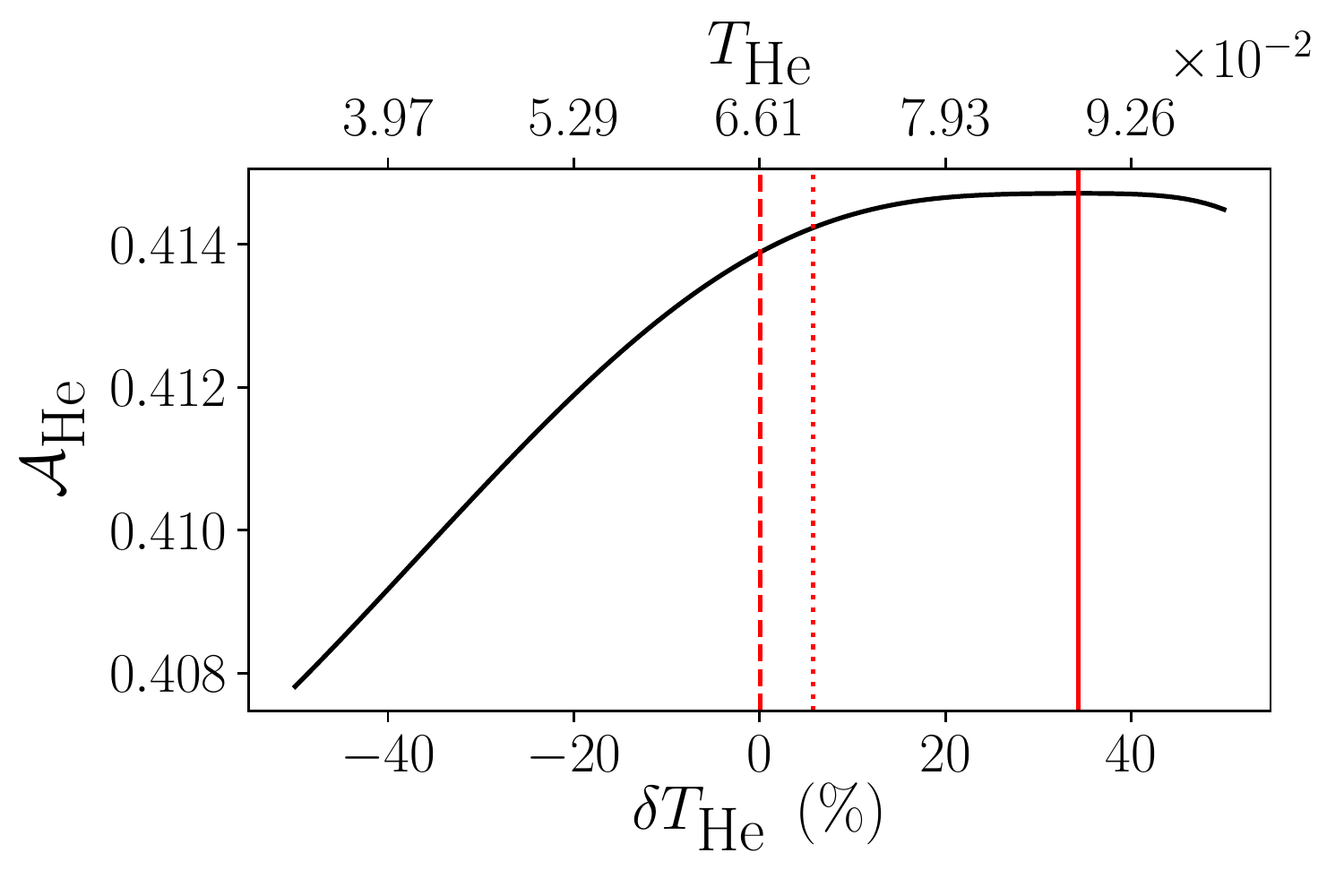}
    \caption{Same as Fig.~\ref{Fig:16CygB-AHe} in KIC8394589's case with the exception that the blue horizontal line corresponding to a one $\sigma$ change in the optimised helium glitch amplitude is not visible as it sits outside of the range of acoustic depth values considered.}
    \label{Fig:8394589-AHe}
\end{figure}

\begin{table}
\caption{Comparison between different estimates of the dimensionless acoustic depth. `model' values corresponds to the ones obtained through the older approach via a stellar model, `fit' values are the results of using the linearly adjusted relation (Eq.~\ref{Eq:TauFit}), and `optimal' values are obtained after the Brent optimisation step, corresponding to the $\chi^2$ minimum.}
\centering
\begin{tabular}{cccc}\label{Tab:THeComp}
Id & model $T_{\textrm{He}}$ & fit $T_{\textrm{He}}$ & optimal $T_{\textrm{He}}$ \\
\hline
16CygA & $0.0756$ & $0.0702$ & $0.0839$ \\
16CygB & $0.0751$ & $0.0717$ & $0.0753$ \\
KIC8394589 & $0.0699$ & $0.0661$ & $0.0888$ \\
KIC8006161 & $0.0731$ & $0.0708$ & $0.0657$ 
\end{tabular}
\end{table}

\subsection{Impact on the inferred helium abundance}\label{Sec:HeEst}
In order to assess the impact of using the optimised value for $T_{\textrm{He}}$ on the inferred helium abundance, we build two stellar models for each of the four targets. The procedure is similar to that presented in \cite{2020A&A...644A..37F}. Considering the same set of reference physical ingredients as their reference models, we use a Levenberg-Marquardt minimisation algorithm to find a CLES model that best matches the four observed seismic indicators: $\Delta$, $\hat{r}_{01}$, $\hat{r}_{02}$ and $\mathcal{A}_{\textrm{He}}$. The free parameters of the procedure are the stellar age (t), mass (M), initial hydrogen abundance ($X_0$), and initial metals ratio ($\left(Z/X\right)_0$).  We consider that a model properly fits the observed data when the theoretical seismic indicators are within one $\sigma$ of their observed value. In other words, a model is considered acceptable when $\chi^2~<=~1$, with
\begin{equation}
    \chi^2 = \sum\limits^L_{i=1} \frac{\left(\theta_{i,\textrm{obs}} - \theta_{i,\textrm{th}}\right)^2}{\sigma_i^2},
\end{equation}
where $\theta_i$ is the i-th of the L constraints, the `obs' and `th' subscript corresponding to observed and theoretical values, respectively, and $\sigma_i$ the uncertainties associated with the indicator.

The only difference between the two models lies in the determination of the helium glitch amplitude. For one set of models, dubbed `old', $T_{\textrm{He}}$ is obtained following \cite{2020A&A...644A..37F}'s approach. That is, we first build a model that adjusts $\Delta$, $\hat{r}_{01}$, and $\hat{r}_{02}$ and compute the integral form in Eq.~\ref{Eq:TauMod} to provide an estimate of $T_{\textrm{He}}$. The amplitude of the helium glitch signature can then be measured and used as a constraint to produce the final model, yielding the helium abundance estimate. The motivation behind this approach was that these three indicators are completely independent of the helium glitch contribution expressed in the \who basis and should allow us to constrain the most part of the stellar structure.
The second estimate, dubbed `new', uses the optimised -- via Brent's optimisation scheme -- $T_{\textrm{He}}$ to retrieve $\mathcal{A}_{\textrm{He}}$ without needing a partial modelling of the star. The provided value is therefore completely model independent. Then, a model representative of the four indicators is built to estimate the helium abundance of the target.

\begin{figure}
    \centering
    \includegraphics[width=\linewidth]{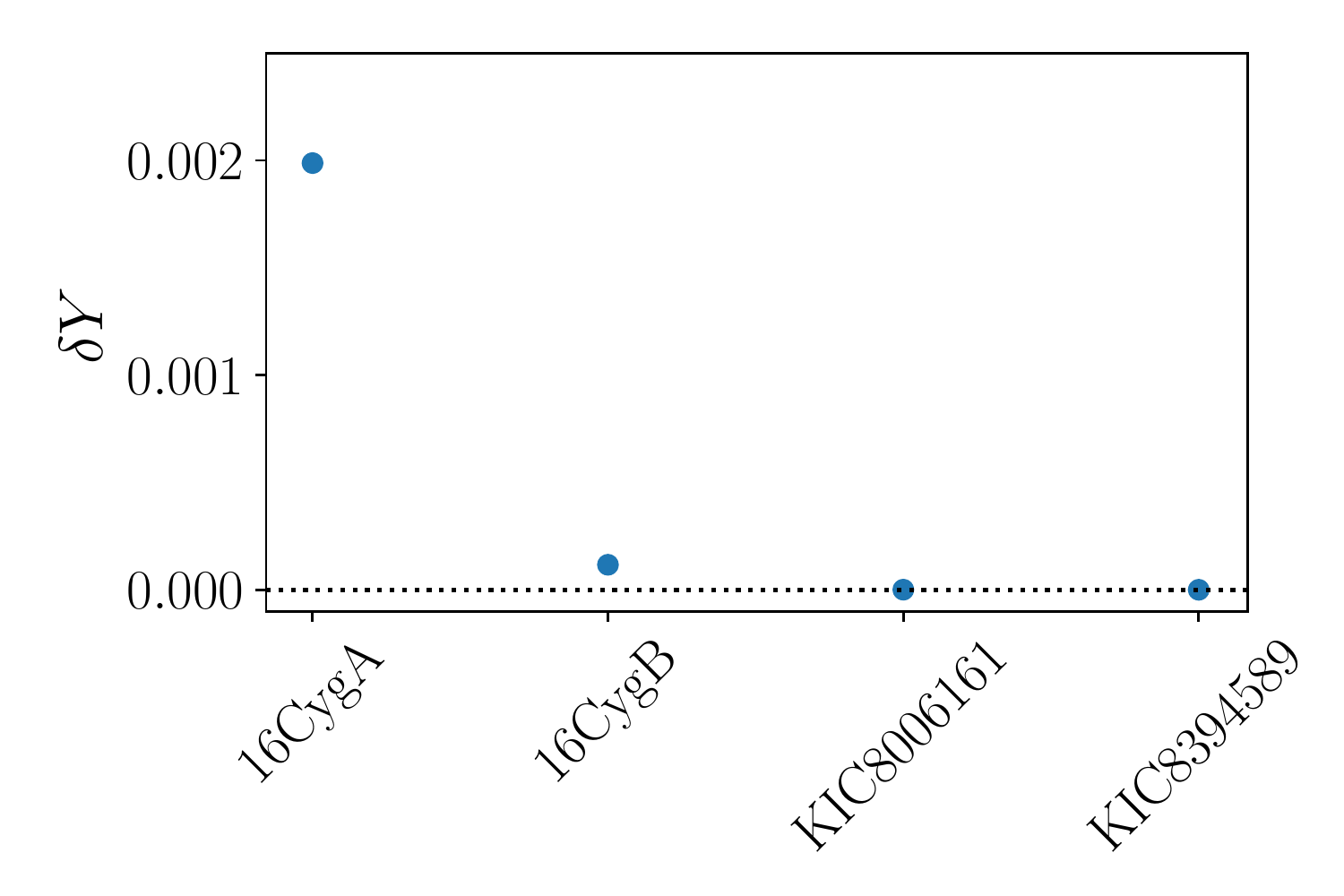}
    \caption{Difference between the two inferred values for the initial helium abundance for the four stars considered in this paper. These two values use either the model value of $T_{\textrm{He}}$ or the Brent optimised one.}
    \label{fig:YComp}
\end{figure}

The results are shown in Fig.~\ref{fig:YComp} where we compute the difference in helium abundances between the `old' and `new' approaches for each of the four stars. We observe that the differences in values obtained with the two approaches, never exceed $0.002~\textrm{dex}$. For comparison, typical uncertainties found in the literature are in the range $\sigma\left( Y_0 \right) \in \left[0.01,0.05\right]$ \citep[for example, either using more sophisticated modelling approaches or acounting for the impact of the elected physical prescription in the stellar models]{2020A&A...644A..37F,2021MNRAS.500...54N,2022MNRAS.515.1492V}. In our opinion, the uncertainties on the initial helium abundance provided by our local model search algorithm are not realistic as they are the product of the inverse Hessian matrix, computed through finite differences. Therefore, these are prone to imprecisions in the derivatives computation and we do not display them. To robustly estimate these uncertainties, more sophisticated modelling approaches (such as MCMC simulations) would be necessary. We also insist on the fact that the uncertainties on the individual seismic indicators used are the result of the propagation of the frequency uncertainties. In all the cases, we observe that the helium abundance estimate has barely changed in comparison to the `old' approach and that the values remain within typical literature uncertainties of a zero difference (for the sake of completeness, stellar parameters for the four stars in the two cases are given in Appendix~\ref{Sec:App-Par}). This is a direct consequence of the robustness of our approach and its relative insensitivity to the acoustic depth parameter, as shown in Sect.~\ref{sec:AHe}. Additionally, we observe that, in some instances, the models computed with the older approach are already compatible with the new measure of $\mathcal{A}_{\textrm{He}}$ and, as a consequence, the optimal model, and inferred helium abundance, has not changed.  
For comparison's sake, we compare the values of the initial helium abundance we infer with the optimised value of $T_{\textrm{He}}$ to the ones obtained by \cite{2019MNRAS.483.4678V} and \cite{2021MNRAS.500...54N}. This is presented in Table~\ref{tab:YComp}. To avoid an artificial spread of the values we considered only the models computed with MESA \citep{2011ApJS..192....3P} in \cite{2019MNRAS.483.4678V}'s study and only the models computed with grid A from \cite{2021MNRAS.500...54N}. This prevents us from factoring in the impact of different physical prescriptions and evolution models within the same studies. We observe that, in all cases, our values agree within one $\sigma$ of the ones retrieved by \cite{2019MNRAS.483.4678V}. Compared to \cite{2021MNRAS.500...54N}'s values, we find that they agree for KIC8006161 and KIC8394589 that have the largest uncertainties in $Y_0$ and that they barely disagree for 16Cyg A and B. This might be due to the fact that \cite{2021MNRAS.500...54N}'s results are on the low side of $Y_0$ values and that they do not account for the helium glitch information in their fitting procedure, they rather use the individual frequencies. Overall, this shows the validity of our refined approach.

\begin{table}
\caption{Comparison of initial helium abundances retrieved with the present approach with literature values in \protect\cite{2019MNRAS.483.4678V} and \protect\cite{2021MNRAS.500...54N}. For consistency, we consider only the values retrieved with the MESA evolution code \protect\citep{2011ApJS..192....3P} in \protect\cite{2019MNRAS.483.4678V}'s case and only results from Grid A of \protect\cite{2021MNRAS.500...54N}.}
\centering
\begin{tabular}{cccc}
Star & Present work & \cite{2019MNRAS.483.4678V} & \cite{2021MNRAS.500...54N} \\
\hline
16CygA & $0.292 \pm 0.015$ & $0.289 \pm 0.013$ & $0.258 \pm 0.014$ \\
16CygB & $0.296 \pm 0.007$ & $0.285 \pm 0.018$ & $0.271 \pm 0.013$ \\
KIC8006161 & $0.282 \pm 0.043$ & $0.259 \pm 0.033$ & $0.265 \pm 0.025$ \\
KIC8394589 & $0.276 \pm 0.026$ & $0.311 \pm 0.023$ & $0.263 \pm 0.025$
\end{tabular}
\label{tab:YComp}
\end{table}

\section{Conclusion}\label{Sec:Con}
The precise measurement of the helium glitch signature is an essential body of work in order to accurately estimate the helium abundance of low mass stars and lift the degeneracy between helium content and mass \citep{2014A&A...569A..21L}, hindering our ability to retrieve precise stellar parameters. Nonetheless, due to its faint nature, its detection is a difficult task and numerous approaches have been proposed \citep[e.g.][]{2014ApJ...782...18M,2014ApJ...790..138V,2019A&A...622A..98F}. While the most important piece of information carried by the glitch is its amplitude, as it correlates with the surface helium content \citep{2004MNRAS.350..277B,2014ApJ...790..138V,2019A&A...622A..98F,2021A&A...655A..85H}, its acoustic depth remains an important parameter as it is necessary to estimate it to then measure the amplitude. Additionally, this parameter appears non-linearly in the glitch's expression. In the present paper, we developed an approach to automatically estimate the helium glitch acoustic depth. With a simple minimisation step, we are able in a fraction of a second to automatically provide a robust and accurate value for the helium glitch acoustic depth, and therefore the helium glitch amplitude. We demonstrated for solar-like stars that our method is robust with respect to the exact definition of the helium glitch acoustic depth. The simple and fast approach we propose leads to measurements of the helium glitch amplitude that are consistent with the older approach implemented with \who \citep{2019A&A...622A..98F}. Using four Kepler LEGACY targets, we further demonstrated that the helium abundance inferred using the revised glitch amplitude is consistent with both the older approach and the values presented in independent studies \citep{2019MNRAS.483.4678V,2021MNRAS.500...54N}. As the precise retrieval of the helium glitch signature is crucial to the accurate measurement of the helium abundance of low-mass stars, the proposed method proves to be an excellent candidate as it allows us to fully automate the \who method by alleviating the need for a partial modelling of the considered target. Thanks to \who's precision and speed of execution, the model independent glitch signature adjustment is automatically carried out in less than a second. Furthermore, as we demonstrated, the model independent approach proposed here is compatible with other studies, offering many advantages. This opens the possibility to robustly analyse very large samples of data as can be expected from future missions such as PLATO. An implementation of the acoustic depth estimation presented in this paper within \who will be made available on \who's GitHub page (\url{https://github.com/Yuglut/WhoSGlAd-python}) by the time of publication.

\section*{Acknowledgements}
M.F. and A.-M.B. acknowledge the support STFC consolidated grant ST/T000252/1.\\
The authors would like to thank the referees for their constructive remarks and suggestions, significantly improving the present paper.

\section*{Data availability}
The data underlying this article are available in the article and in its online supplementary material.

\bibliographystyle{mnras}
\bibliography{bibli}

\begin{appendix}

\section{Additional $\chi^2$ and $A_{\textrm{He}}$ landscapes}\label{Sec:app-chi}
\begin{figure}
    \centering
    \includegraphics[width=\linewidth]{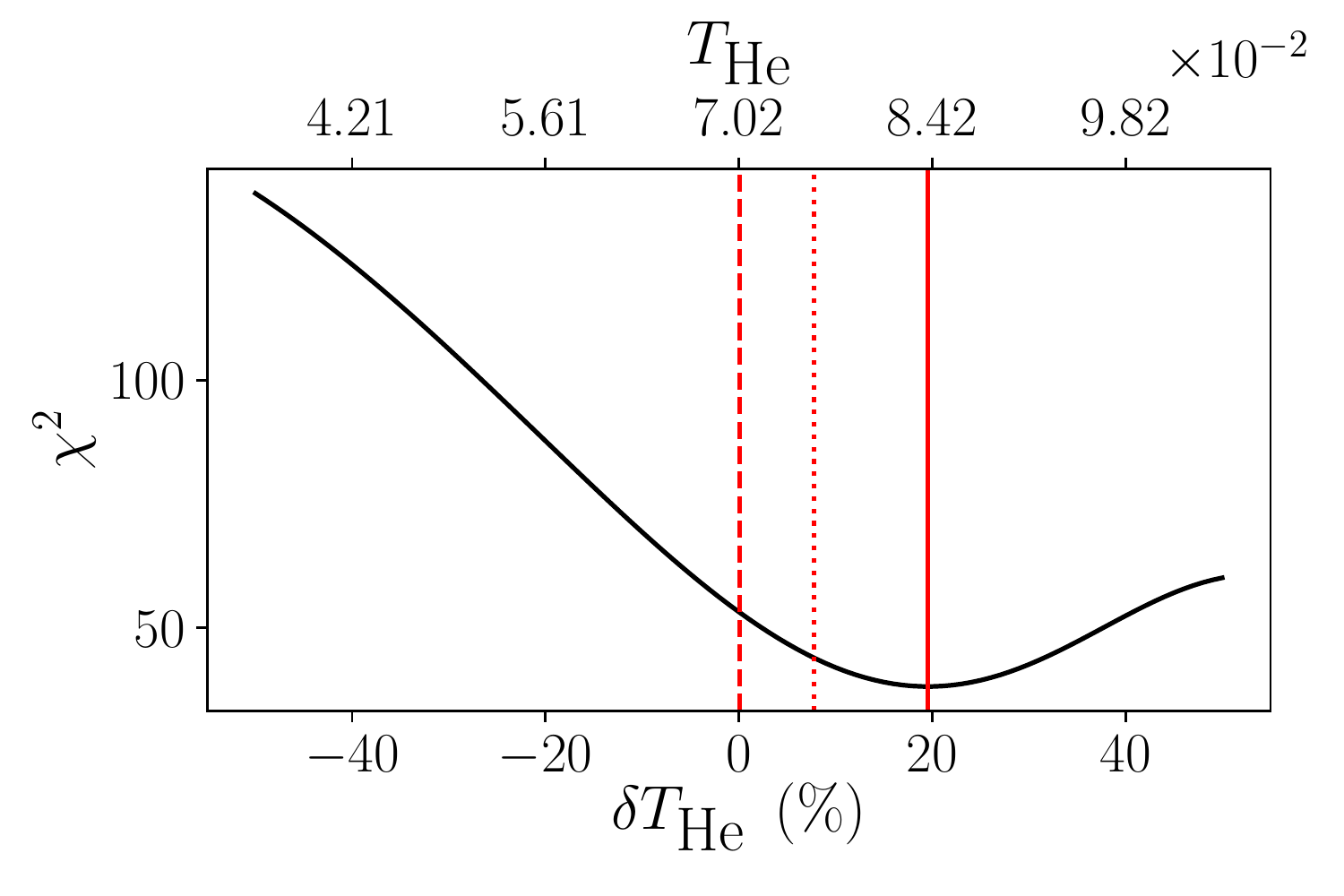}
    \caption{Evolution of the agreement between fitted and observed frequencies of 16CygB, expressed as a $\chi^2$ value, as a function of the helium glitch acoustic depth, expressed as the relative difference with respect to the linear estimate (in $\%$). The $\chi^2$ minimum, obtained via Brent's minimisation procedure, is represented by the red vertical line. We also show the linear estimate as a vertical dashed line and the value obtained with the `old' approach as a dotted one.}
    \label{Fig:16CygA-chi}
\end{figure}

\begin{figure}
    \centering
    \includegraphics[width=\linewidth]{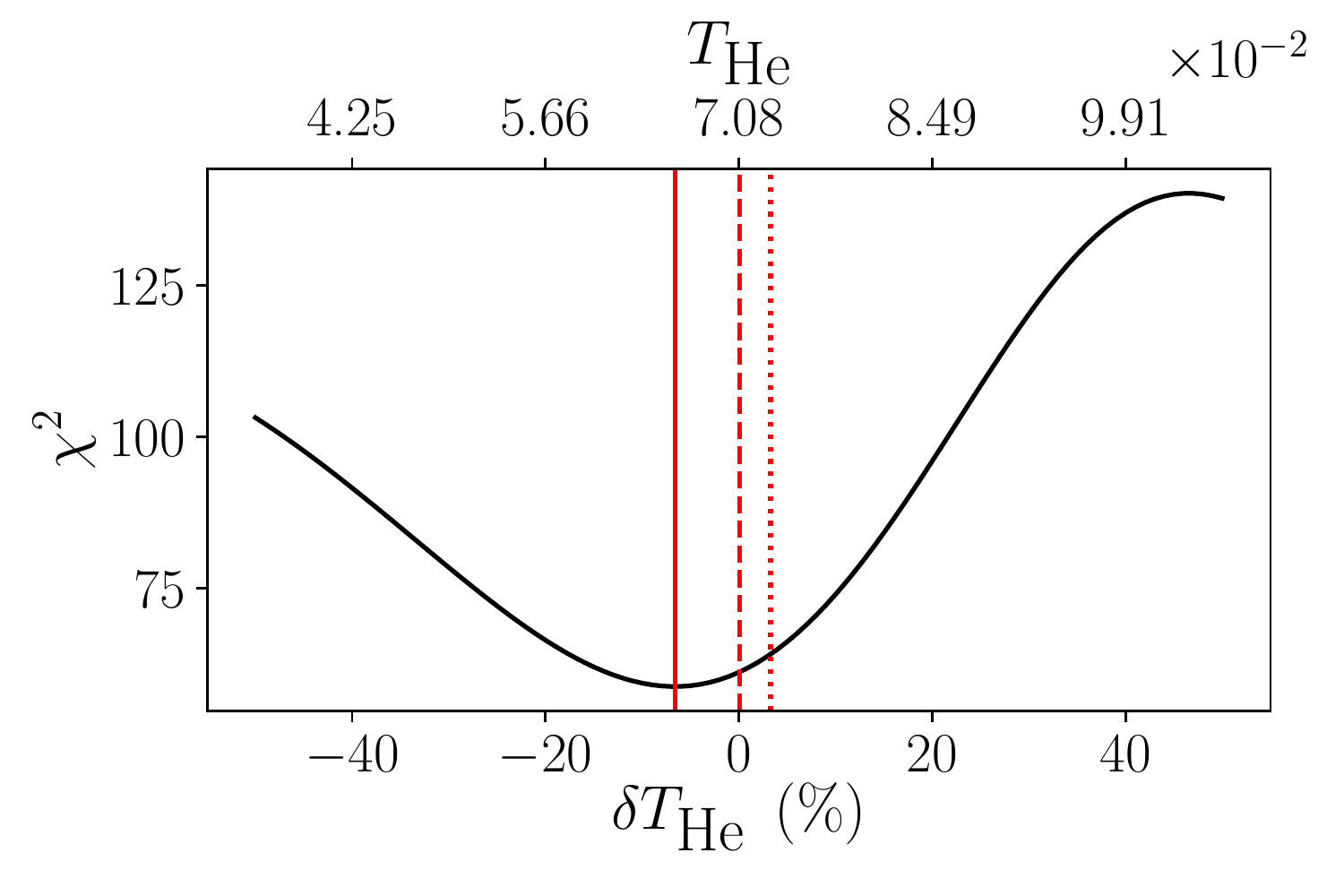}
    \caption{Same as Fig.~\ref{Fig:16CygA-chi} in KIC8006161's case.}
    \label{Fig:8006161-chi}
\end{figure}

\begin{figure}
    \centering
    \includegraphics[width=\linewidth]{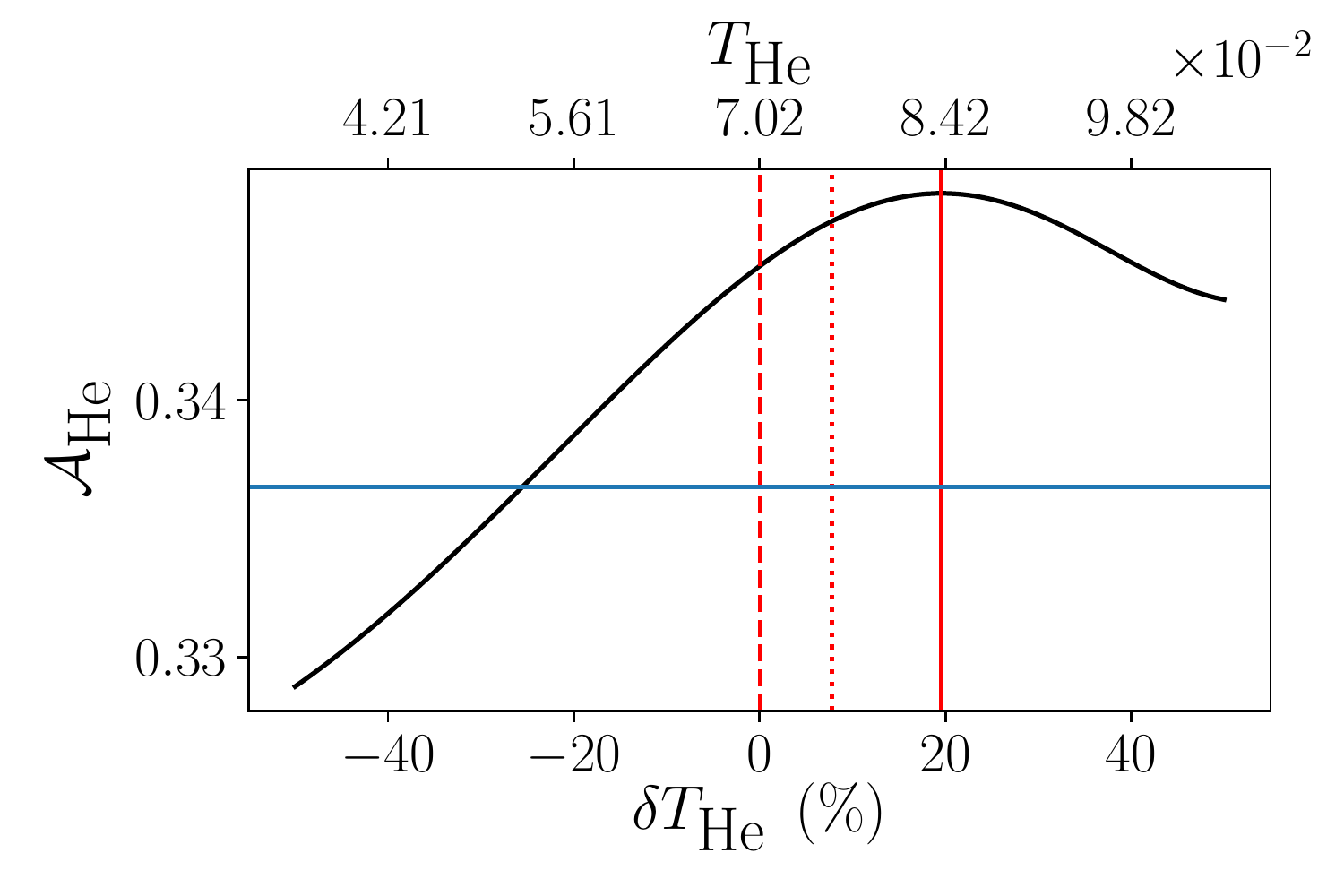}
    \caption{Evolution of the helium glitch amplitude of 16CygA with the variation of the helium glitch acoustic depth with respect to the linear estimate (Eq.~\ref{Eq:TauFit}). This variation is expressed as a relative difference in percent. The $\chi^2$ minimum, obtained via Brent's minimisation procedure, is represented by the red vertical line. We also show the linear estimate as a vertical dashed line and the value obtained with the `old' approach as a dotted one. The blue horizontal line corresponds to a $1~\sigma$ variation in amplitude from the optimal value.}
    \label{Fig:16CygA-AHe}
\end{figure}

\begin{figure}
    \centering
    \includegraphics[width=\linewidth]{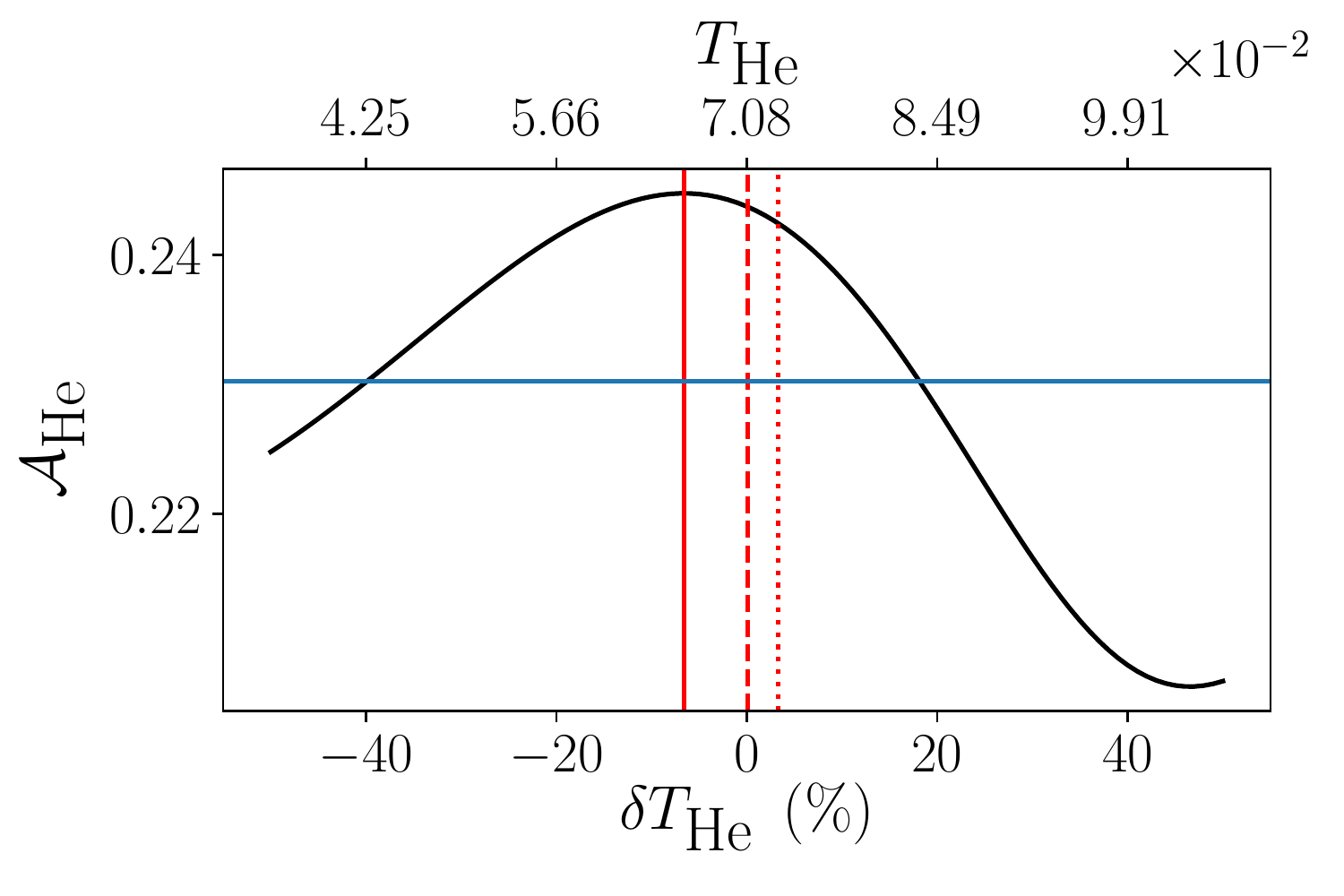}
    \caption{Same as Fig.~\ref{Fig:16CygA-AHe} in KIC8006161's case.}
    \label{Fig:8006161-AHe}
\end{figure}

We provide in Figs.~\ref{Fig:16CygA-chi} and \ref{Fig:8006161-chi} the $\chi^2$ landscapes for 16CygA and KIC8006161 and in Figs.~\ref{Fig:16CygA-AHe} and \ref{Fig:8006161-AHe} their $A_{\textrm{He}}$ landscapes.

\section{Modelled stellar parameters}\label{Sec:App-Par}
In this section, we give the stellar parameters computed for the four stars considered in this paper, using either the `old' approach or the one developed in this paper. We insist on the fact that the quoted uncertainties are the result of the Hessian matrix inversion (during the Levenberg-Marquardt procedure) and is highly prone to imprecision in the derivatives. Therefore, these only serve to give an order of magnitude to the confidence interval.

\begin{table}[]
    \centering
    \caption{Comparison of the optimised parameters using either the `old' or the revised method for the four stars considered inm this paper.}
    \begin{tabular}{cccc}
        & & Old & New \\ \hline
                & $M~\left(M_{\odot} \right)$ & $1.057 \pm 0.025$ & $1.056 \pm 0.023$ \\
                & $t~\left(\textrm{Gyrs}\right)$ & $6.80 \pm 0.18$ & $6.78 \pm 0.12$ \\
        16Cyg A & $X_0$ & $0.684 \pm 0.015$ & $0.687 \pm 0.008$ \\
                & $\left(Z/X\right)_0$ & $0.035 \pm 0.001$ & $0.034 \pm 0.002$ \\
                & $Y_0$ & $0.292 \pm 0.015$ & $0.290 \pm 0.009$  \\ \hline
                & $M~\left(M_{\odot} \right)$ & $1.011 \pm 0.008$ & $1.014 \pm 0.007$ \\
                & $t~\left(\textrm{Gyrs}\right)$ & $6.96 \pm 0.08$ & $6.96 \pm 0.09$ \\
        16Cyg B & $X_0$ & $0.679 \pm 0.007$ & $0.678 \pm 0.006$ \\
                & $\left(Z/X\right)_0$ & $0.037 \pm 0.002$ & $0.038 \pm 0.002$ \\
                & $Y_0$ & $0.296 \pm 0.007$ & $0.296 \pm 0.006$   \\ \hline
                & $M~\left(M_{\odot} \right)$ & $0.979 \pm 0.118$ & $0.979 \pm 0.118$ \\
                & $t~\left(\textrm{Gyrs}\right)$ & $5.50 \pm 0.56$ & $5.50 \pm 0.56$ \\
     KIC8006161 & $X_0$ & $0.685 \pm 0.031$ & $0.683 \pm 0.031$ \\
                & $\left(Z/X\right)_0$ & $0.050 \pm 0.040$ & $0.050 \pm 0.039$ \\
                & $Y_0$ & $0.282 \pm 0.043$ & $0.282 \pm 0.043$   \\ \hline
                & $M~\left(M_{\odot} \right)$ & $1.085 \pm 0.024$ & $1.085 \pm 0.024$ \\
                & $t~\left(\textrm{Gyrs}\right)$ & $3.53 \pm 0.14$ & $3.53 \pm 0.14$ \\
     KIC8394589 & $X_0$ & $0.711 \pm 0.025$ & $0.711 \pm 0.025$ \\
                & $\left(Z/X\right)_0$ & $0.018 \pm 0.006$ & $0.018 \pm 0.006$ \\
                & $Y_0$ & $0.276 \pm 0.026$ & $0.276 \pm 0.026$ 
    \end{tabular}
    \label{tab:ModPar}
\end{table}


\end{appendix}

\bsp	
\label{lastpage}
\end{document}